# Molecular-Kinetic Simulations of Escape from the Ex-planet and Exoplanets: Criterion for Transonic Flow

Robert E. Johnson[1,2], Alexey N. Volkov[1,3] and Justin T. Erwin[1,4]

[1] Engineering Physics, University of Virginia, Charlottesville, VA 22904-4745, USA
[2] Physics Department, New York University, NY 10003-6621, USA
[3] Department of Mechanical Engineering, University of Alabama, Tuscaloosa, AL 35487
[4] Department of Planetary Science, University of Arizona, Tucson, AZ 85721, USA

**Abstract:**
The equations of gas dynamics are extensively used to describe atmospheric loss from solar system bodies and exoplanets even though the boundary conditions at infinity are not uniquely defined. Using molecular-kinetic simulations that correctly treat the transition from the continuum to the rarefied region, we confirm that the energy-limited escape approximation is valid when adiabatic expansion is the dominant cooling process. However, this does not imply that the outflow goes sonic. In fact in the sonic regime, the energy limited approximation can significantly under estimate the escape rate. Rather large escape rates and concomitant adiabatic cooling can produce atmospheres with subsonic flow that are highly extended. Since this affects the heating rate of the upper atmosphere and the interaction with external fields and plasmas, we give a criterion for estimating when the outflow goes transonic in the continuum region. This is applied to early terrestrial atmospheres, exoplanet atmospheres, and the atmosphere of the ex-planet, Pluto, all of which have large escape rates. The paper and its erratum, combined here, are published: ApJL 768, L4 (2013); ApJ, 779, L30 (2013).

## Introduction

Rapid atmospheric escape is often described as a gas that goes sonic, sometimes called blow-off (Hunten 1982), a process that can account for certain isotope ratios on terrestrial planets. Transonic models have also been used to describe rapid escape from exoplanets (e.g., Murray-Clay et al. 2009) and from Pluto (e.g., Strobel 2008). However, we recently showed that this model for Pluto gave an incorrect upper atmospheric structure (Tucker et al. 2012; Erwin et al. 2013).

In simulating rapid escape using continuum gas dynamics, the Jeans expressions at the exobase have been applied for the uncertain boundary conditions at infinity (e.g., Tian et al. 2008). More often, a sonic point is assumed to occur at some altitude, above which the density and temperature dependence can be simply characterized (Parker 1964a,b). The so-called energy-limited escape rate, extensively applied to exoplanet atmospheres (e.g., Lammer et al. 2009), is often assumed to imply that sonic boundary conditions are applicable (e.g., Erkaev et al. 2012). Here we use molecular-kinetic simulations to show that is not the case.

We briefly review the continuum and molecular-kinetic models, and then present results of our simulations. These test the applicability of the energy-limited escape rate and our proposed criterion for determining whether sonic or kinetic upper boundary conditions are applicable. The results are applied to escape from Pluto, early terrestrial planet, and exoplanet atmospheres.

**Models**

We describe escape from a one-dimensional (1D), steady state, single component atmosphere as illustrative, leaving out thermal transport by horizontal flow. For radial distance $r$, flow speed $u$, number density $n$, temperature $T$, pressure $p = nkT$, and escape rate $\Phi = 4\pi r^2 n u = $ const, the momentum and energy equations are often used ignoring viscosity:

$$n \frac{d}{dr}(mu^2/2 - U) = -\frac{dp}{dr} \qquad (1)$$

$$\frac{d}{dr}\left[\Phi(mu^2/2 + C_p kT - U) - 4\pi r^2 \kappa \frac{dT}{dr}\right] = 4\pi r^2 n \, q_a(r) \qquad (2)$$

Here $k$ is the Boltzmann constant, $\kappa = \kappa(T)$ the thermal conductivity, $C_p$ the specific heat at constant pressure, $m$ the molecular mass, $U = U(r) = GMm/r$ the gravitational energy ($G$ the gravitational constant, $M$ the planet's mass), and $q_a(r)$ is the *net* heating rate per molecule produced by incident photons or plasma particles, in which we include radiative cooling. Knowing the density, $n_0$, and temperature, $T_0$, at a lower boundary, $r = r_0$, a unique solution requires two other conditions, typically at the upper boundary. The gravitational energy is characterized by the Jeans parameter, $\lambda(r) = U/kT$ and the rarefaction by the Knudsen number, $Kn(r) = l_c/H$, the ratio of the mean free path of gas molecules, $l_c$, to the local scale height, $H = -n/(dn/dr)$. For an escaping gas at large distances from the source, where free molecular flow occurs, $H \to r/2$; in the hydrostatic regime $H \to r/\lambda(r)$.

The Jeans expressions for the number, $\Phi_J$, and thermal, $\langle E\Phi \rangle_J$, escape rates have been used as upper boundary conditions for Equations (1-2):

$$\Phi_J = 4\pi r_x^2 n_x \sqrt{\frac{kT_x}{2\pi m}}(1 + \lambda_x)\exp(-\lambda_x) \qquad (3a)$$

$$\langle E\Phi \rangle_J = kT_x \Phi_J \left(\frac{1}{1+\lambda_x} + C_p - \frac{3}{2}\right) \qquad (3b)$$

The subscript 'x' indicates quantities evaluated at the nominal exobase, $r = r_x$, where $Kn(r_x) \approx 1$, often assumed to be the upper boundary of the continuum region. When the upper atmosphere heating rate is large, the equations are more often solved through a sonic point, $r = r_*$, where $u_* = c$ ($c = \sqrt{\gamma kT/m}$ the sound speed, $\gamma = C_p/C_V$, $C_V$ the specific heat at constant volume). For $r \gg r_*$, $n$ and $T$ decay as power laws (Parker 1964a,b). Unfortunately those continuum solutions for which Jeans escape is applicable and those for which a sonic point is reached in the continuum region do not simply track from one to the other as the heating rate increases.

Kinetic models *can* simulate both continuum ($l_c \ll H$) and non-continuum (transitional, $l_c \sim H$, and free molecular, $l_c \gg H$) gas flows and can, therefore, describe the change from Jeans-like to transonic escape. Since such simulations track particles in the potential well of the body (or bodies) of interest, escape is a natural outcome. We numerically implement a kinetic description of rarefied gas flow in an upper atmosphere based on the Boltzmann kinetic equation using the direct simulation Monte Carlo

(DSMC) method (Bird 1994). In this method, the gas flow is represented by a large set of representative atoms or molecules that are tracked subject to binary collisions and gravity (Volkov 2011a,b). Heating of the atmosphere is implemented by scaling the thermal velocities of the representative molecules according to the local energy deposition rate.

The lower boundary of the simulation region, $r = r_0$, is below the depth at which the UV/EUV or plasma energy deposition occurs and $Kn(r_0) \ll 1$. Because the density drops rapidly with increasing $r$, but escape occurs at large $r$ where the density is low, DSMC simulations starting at small $Kn(r_0)$ can require an enormous number of particles to accurately describe escape. Therefore, we also use a hybrid continuum/kinetic model (Tucker et al. 2012; Erwin et al. 2013) in which the Equations (1-2) are solved at $Kn(r) < \sim 0.1 - 0.01$, where the gas is collisionally dominated, and the velocity and internal energy distributions are reasonably well represented by Maxwellians, and then iteratively couple it to a DSMC simulation in the rarefied region.

**Heating**

Parker (1964a,b) used Equations (1-2) to describe escape when the dominant heat source is internal, as it is for expansion of the solar corona: i.e., $q_a(r) = 0$ for $r > r_0$. This model was subsequently applied to planetary atmospheres primarily heated at $r < r_0$. For Jeans parameters at $r = r_0$ as large as $\lambda(r_0) = \lambda_0 \sim 40$, such models were assumed to produce a transonic expansion, often referred to as 'slow hydrodynamic escape' (e.g., Strobel 2008). Although, rapid escape *can* occur for relatively large $\lambda_0$ and $Kn(r_0) \ll 1$, for $\lambda_0 > \sim (C_p + \gamma/2)$ the gas does not go sonic in the collision-dominated region and the escape rate is a factor of a few larger than the Jeans rate (Volkov et al. 2011a,b). Ignoring the thermal conductivity in Equation (2), this corresponds to the enthalpy of fluid particles becoming sufficient to produce a transonic, isentropic expansion starting at $r_0$. For smaller $\lambda_0$ a non-equilibrium region, called a Knudsen layer, forms above $r = r_0$ and there is a steep transition with decreasing $\lambda_0$ to supersonic escape at $\lambda_0 \sim 2.1$ and $\sim 2.8$-$3.5$ for monatomic and diatomic gases respectively (Volkov & Johnson 2013).

In an upper atmosphere heated by short wavelength radiation, or by incident plasma particles, escape is driven by the energy absorbed. Energy absorbed below $Kn(r) \sim 0.1$ is typically converted to heat using an efficiency, $\varepsilon$, that depends on the radiation type and atmospheric composition. The heating rate is either directly calculated or a value of $\varepsilon$ is estimated: often a constant ($\sim 0.15 - 0.4$) up to the exobase where it goes to zero. Typically the gas-dynamic equations for exoplanet or early terrestrial atmospheres are then solved with Jeans-like or sonic upper boundary conditions. As discussed below, we used a hybrid continuum/DSMC model to describe escape from Pluto heated by the solar UV/EUV and a DSMC model to describe escape from an atmosphere in which, the heating is assumed to occur in a narrow layer.

**Energy Limited Escape**

Because thermal conduction in the upper atmosphere is inefficient, adiabatic cooling by escape or horizontal transport often dominates (e.g., Erwin et al. 2013). For a *globally averaged* heating rate and adiabatic cooling, integration of Equation (2) gives a rough upper bound to the escape rate, $\Phi_{EL}$ (e.g., Lammer et al. 2009):

$$\Phi_{EL} \approx Q_{net}/(U - C_p kT - mu^2/2)|_{r_u}^{r_0} . \tag{4}$$

where the denominator is the difference in the average energy of a molecule between the lower, $r_0$, and upper, $r_u$, boundaries of the simulation. In Equation (4) $r_0$ is assumed to be below the heated region, above which adiabatic cooling dominates; $Q_{net} = 4\pi \int_{r_0}^{\infty} r^2 q_a(r) dr$ is the integrated heating + radiative cooling rate. This expression is often referred to as the energy-limited rate, although Watson et al. (1981) discussed a related quantity. For $r_0$ deep in the gravitational well and $\lambda_0 \gg C_p$, Equation (4) is often approximated as

$$\Phi_{EL} \approx Q_{net}/U(r_0) . \tag{5}$$

Assuming a small fraction of $Q_{net}$ is deposited for $Kn(r) > \sim 0.1$, so non-thermal escape processes (Johnson et al. 2008) can be ignored, we showed that UV/EUV heating of Pluto's atmosphere resulted in an escape rate very close to that in Equation (5), *but* the gas did not go sonic below the exobase. Rather, a large expansion of the upper atmosphere occurred (Tucker et al. 2012; Erwin et al. 2013). Similarly, Tian et al. (2008) found that above a heating threshold, the atmosphere rapidly expanded and the escape rate increased with increasing EUV heating, consistent with energy-limited escape, even though the gas remained subsonic. Therefore, the energy limited escape rate is *not* contingent on the flow going sonic below the exobase. If the heating rate is increased to the point where the atmosphere does go sonic in the continuum region, energy limited escape can still apply if one accounts for the large $u$, enhanced radiative cooling, recombination in an ionized atmosphere, etc. (e.g., Murray-Clay et al. 2009).

**Criteria for Transonic Solutions**

Since the isentropic approximation and energy-limited escape are applicable to both sub-sonic and transonic rapid outflows, we use the Mach number, $Ma = u/\sqrt{\gamma kT/m}$, and rewrite Equation (5):

$$Q_{net} \approx 4\pi r^2 n \, Ma \sqrt{\frac{\gamma}{\lambda} \frac{U(r)}{m}} U(r_0) \tag{6}$$

Because the boundary conditions for sub-sonic and transonic solutions differ, it is important to be able to estimate the minimum value of $Q_{net}$ required to apply sonic boundary conditions, which we will call $Q_c$. Assuming the sonic point, $r = r_*$, occurs in the continuum region, the flow can be effectively approximated by the isentropic model:

$$2mc^2(r_*) = U(r_*) + (\gamma - 1) \frac{r_* q(r_*)}{u(r_*)}. \tag{7}$$

(e.g., Murray-Clay et al. 2009). For $q(r_*) \approx 0$ Equation (7) reduces to $\lambda(r_*) = \lambda_* \approx 2\gamma$ and transonic escape occurs when $Ma > 1$ in Equation (6) giving:

$$Q_{net} > Q_c \approx 4\pi r_*^2 n_* \sqrt{\frac{U(r_*)}{2m}} U(r_0), \tag{8}$$

with $n_* = n(r_*)$. Below we estimate $Q_c$ and test it against molecular kinetic simulations.

We first consider a narrow heating layer located at $r = r_a > r_0$ with $Kn(r_a) \ll 1$, as in Watson et al. (1981) and McNutt (1989), and give an approximate analytic solution to Equations (1-2) in the Appendix. For this case, we performed DSMC simulations for a monatomic gas of hard spheres with $\lambda_0 = 10$, $Kn(r_0) = 10^{-3}$, $r_a/r_0 = 1.1$. Although the results can be roughly scaled, the simulation results shown in Figures 1 and 2 are for an N$_2$ atmosphere with hard sphere collision diameter $4.76 \times 10^{-10} m$, planet mass $M = 4.45 \times 10^{21} kg$, $r_0 = 10^6 m$, $n_0 = 0.993 \times 10^{16}/m^3$, and $T_0 = 100K$. For $Q_{net} = 0$, we showed earlier that these conditions correspond to enhanced Jeans-like escape with $\Phi \approx 1.6\Phi_J$ (Volkov et al. 2011a,b) which for our simulation parameters is $\sim 0.7 \times 10^{-4}\Phi_0$, where $\Phi_0$ is the upward flow across the lower boundary of the simulation, $r_0$: $\Phi_0 = 4\pi r_0^2 n_0 (kT_0/2\pi m)^{0.5}$.

Increasing $Q_{net}$ to find when the outflow goes sonic in the continuum regime, it is seen in Figure 1 that for $Q_{net} \gg 0$ a non-equilibrium layer forms near $r_a$ in which the parallel and perpendicular components of temperature differ. For small $Kn(r_a)$ the flow properties in this layer are analogous to those in the Knudsen layer discussed above for heating below $r_0$. Near $r_a$ the density and temperature change dramatically but the pressure from $r_0$ to $\sim r_a$ can be estimated from the hydrostatic approximation. When the outflow goes sonic, a pressure drop occurs from $\sim r_a$ to $r_*$: $P_c = p_*/p_a$ (Volkov and Johnson 2013). Since the transition layer is narrow, $r_* \approx r_a$, we rewrite Equation (8) using values at $r_0$:

$$Q_{net} > Q_c \approx \langle E\Phi \rangle_0 \left[ \gamma \sqrt{\pi \lambda_0} \left(\frac{r_a}{r_0}\right)^{\frac{5}{2}} P_c \frac{p_a}{p_0} \right] \qquad (9)$$

Here $<E\Phi>_0 = (2kT_0)\Phi_0 = 4\pi r_0^2 n_0 kT_0 (2kT_0/\pi m)^{0.5}$ is the upward Maxwellian energy flux of molecules leaving the source at $r = r_0$. When the effects of thermal conduction are small relative to adiabatic cooling, then Equation (9) can be derived from the analytic expression in the Appendix. The barometric equation gives $p_a/p_0 \approx \exp[\lambda_0(r_0/r_a - 1)]$ with $P_c$ depending on the number of degrees of freedom: $\sim 0.4$ for a monatomic gas. From Figure 1, the transition to supersonic flow occurs for $0.46 < Q_{net}/Q_c < 0.67$ with $Q_c \sim 4.5 \times 10^{10} W$ in Equation 9 for our simulation parameters. Due to the above approximations, Equation (9) overestimates $Q_c$ by about a factor of two. It is also seen in Figure 1(c) that $n(r)$ for the subsonic solution increases slowly for $r \gg r_a$, resulting in a significantly expanded atmosphere, but $n(r)$ for the transonic solution in Figure 1(e) roughly decreases as a power law consistent with transonic escape.

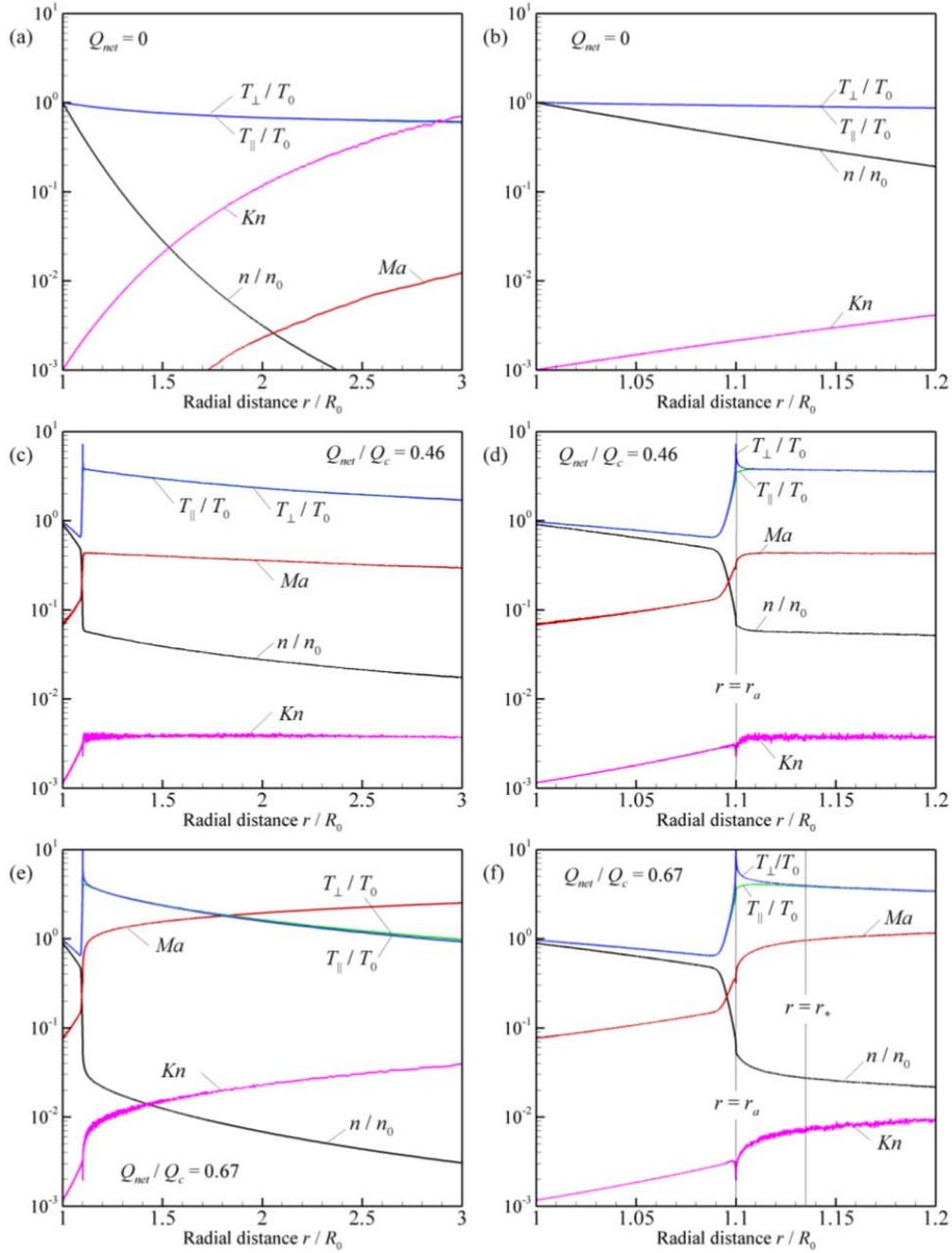

**Figure 1.** DSMC simulations: a monatomic gas of hard spheres at $\lambda_0 = 10$, $Kn(r_0) = 10^{-3}$ heated at $r_a/r_0 = 1.1$, with $r_0 = R_0$. Scaled density, $n/n_0$ (black), parallel, $T_{\parallel}/T_0$ (green), and perpendicular, $T_{\perp}/T_0$ (blue), temperatures, and local Mach (red) and Knudsen (magenta) numbers for $Q_{net}/Q_c = 0$ (a,b; Jean-like escape), $= 0.46$ (c,d; subsonic; $\Phi/\Phi_{EL} = 0.87$), and $= 0.67$ (e,f; transonic at $r_*/r_0 = 1.13$; $\Phi/\Phi_{EL} = 0.65$). Lines: Heated and sonic surfaces.

As important, the escape rate increases dramatically above $Q_{net} = 0$ and becomes close to $\Phi_{EL}$ in Equation (5) in the *subsonic* regime as seen in Figure 2. Although $\Phi$ does not change significantly in the transition to transonic escape, a steep increase *is* seen in the thermal + flow energy removed, $<E\Phi>$. It is also seen that additional heating primarily increases the average kinetic energy of escaping molecules so that Equation (5) becomes a poor approximation if $Q_{net} \gg Q_c$. For very large $Q_{net}$ in Figure 2 the escape rate is limited to a fraction of $\Phi_0$ and $<E\Phi>$ approaches $Q_{net}$. For small $Q_{net}$ downward thermal conduction affects the escape rate (Erwin et al. 2013), which is eventually driven by the temperature at $r_0$.

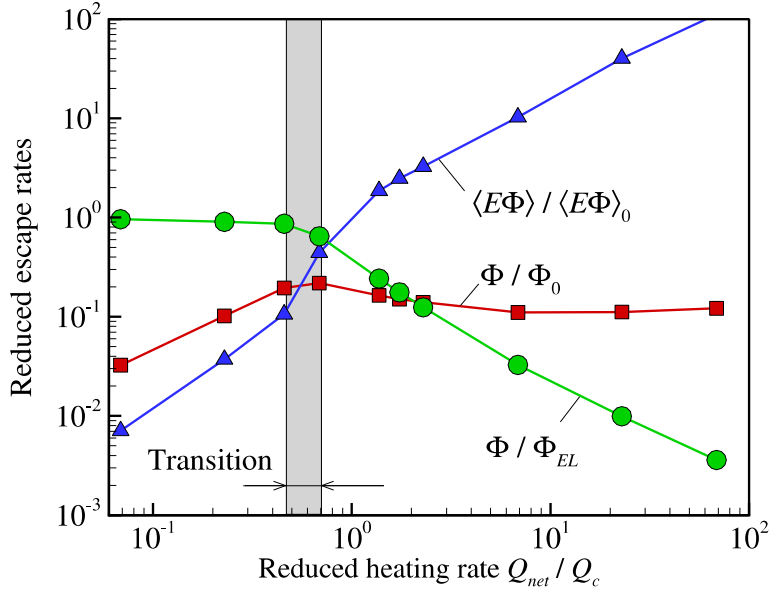

**Figure 2.** Number, $\Phi$ [green circles, red squares] and energy, $\langle E\Phi \rangle$ [blue triangles] escape rates vs. $Q_{net}/Q_C$ calculated as for Figure 1: Scaled to $\Phi_{EL}$ from Equation (5) and also to $\Phi_0$, the upward flow across the lower boundary of the simulation, $r_0$, $\Phi_0 = 4\pi r_0^2 n_0 (kT_0/2\pi m)^{0.5}$ (in Volkov et al. (2011a,b) the symbol $\Phi_{0,0}$ is used). $<E\Phi>$ is scaled to the energy flux of molecules across $r_0$, $<E\Phi>_0 = (2kT_0)\Phi_0 = 4\pi r_0^2 n_0 kT_0 (2kT_0/\pi m)^{0.5}$. Rectangle indicates transition from subsonic to supersonic flow below the exobase. For these simulations $\Phi_0 = 8.6 \times 10^{30} s^{-1}$; $<E\Phi>_0 = 2.4 \times 10^{10} W$; $Q_c \sim 4.5 \times 10^{10} W$ from Equation 9.

We now consider more realistic heating profiles. In order to apply sonic boundary conditions in Equations (1-2), $Kn(r_*)$ must be in the continuum region below some maximum, $Kn_m$: i.e., $Kn(r_*) = l_{c*}/H_* < Kn_m$. Using $H_* \sim r_*/\lambda_*$ and $l_{c*} = 1/(c_c \sigma_c n_*)$, where $c_c$ is determined by the energy dependence of the total collision cross section $\sigma_c$ (e.g., $c_c = \sqrt{2}$, $\sigma_c = \pi d^2$ for the gas of hard sphere molecules of diameter *d*), then $2\gamma/(c_c r_* \sigma_c n_*) < Kn_m$. From Equation (8), we estimate $Q_c$ requiring as that $r_*$ occurs in the continuum domain:

$$Q_{net} > Q_c \approx 4\pi r_* \frac{\gamma}{c_c \sigma_c Kn_m} \sqrt{\frac{2U(r_*)}{m}} U(r_0), \qquad (10)$$

where $r_0 < r_* < r_x$. When there is a sharp change in the gas properties, as for the heated layers discussed above, then $Kn_m < \sim 0.1$ (Volkov and Johnson 2013). However, if the heat is primarily absorbed over a board range of $r$ below $r_x$, then $Kn_m \sim 1$ is sufficient.

It is seen in Equation 10 that $Q_c$ does not *explicitly* depend on $T_0$, consistent with simulations when $\Phi$ is large. Because $Q_c$ depends on the sonic point only via $(r_*)^{1/2}$, a rough lower bound can be obtained by replacing $r_*$ with the mean energy absorption depth, $r_a$ estimated from the absorption cross section, $\sigma_a$. More accurately, at threshold the sonic point approaches $r_x$ so that $r_* \sim r_a[1 + (\sigma_a/c_c\sigma_c) \lambda_{ave}]$ where $\lambda_{ave} \sim (\lambda_a + 2\gamma)/2$ slightly increasing $Q_c$. For a close-in exoplanet, tidal heating can be included in $U(r)$ and ion escape can dominate so that $\sigma_c$ becomes large due to ion-neutral or ion-ion collisions reducing $Q_c$.

For solar minimum, medium, and maximum conditions we simulated Pluto's upper atmosphere at the New Horizons encounter using our hybrid fluid/kinetic model ignoring the interaction with the solar wind and Charon, as well as non-thermal escape (Erwin et al. 2013). For all three cases ($Q_{net}$ = 0.38, 0.78, ~1.6x10$^8$ W) the atmosphere became *highly* extended, but the flow remained subsonic *contrary to all* earlier models (e.g., Strobel 2008) with the escape rate close to the energy-limited estimate in Equation (4). It is seen in Figure 3 that at solar medium $r_x$ is more than twice that obtained when sonic boundary conditions are applied and, although the escape rate is large, using the Jeans boundary conditions results a much better approximation to the upper atmosphere structure.

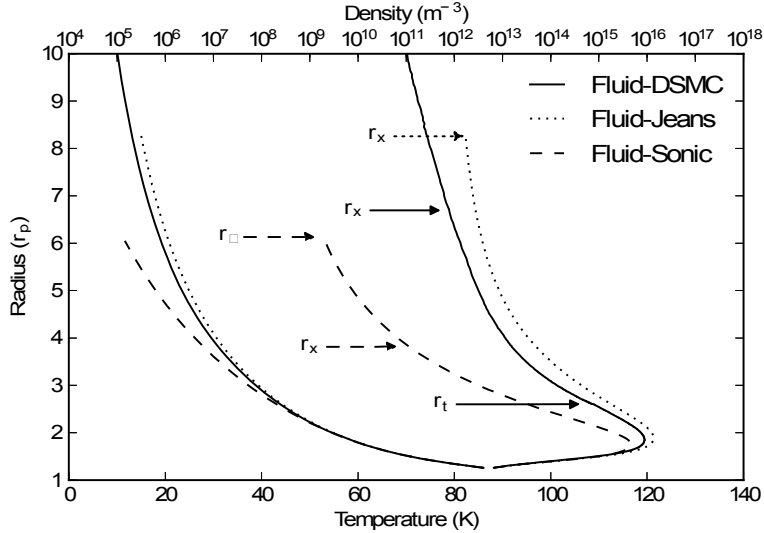

**Figure 3.** Fluid simulations of Pluto's N$_2$ atmosphere at 32Au at solar-medium; upper bc: (solid) fluid/DSMC coupled at $r_t$, $Kn$ = 0.1; $\Phi$ = 2.6x10$^{27}$; (dashed) transonic assumption: $r_* > r_x$, $\Phi$ = 2.5x10$^{27}$/s (Strobel 2008); (dotted) Jeans bc from Equations (3a,b) at $r_x \sim (7-8)r_p$; $\Phi$ = 2.6x10$^{27}$/s; $r_p$ = 1153 km Pluto's radius; $r_0$ = 1.25$r_p$ roughly the visible extinction radius; parameters in VHS-LB model (Erwin et al. 2013): $m$ = 28 amu, $n_0$ = 4x10$^{12}$ cm$^{-3}$, $T_0$ = 88.2 K; $U(r_0)$ = 2.8x10$^{-13}$ ergs ; $\lambda_0$ = 23; $\gamma$ = 7/5 ; $\sigma_c \sim$9x10$^{-15}$ cm$^2$; Kn($r_0$) ~ 10$^{-6}$.

For UV/EUV absorption at $r_a \sim$1.5 times Pluto's radius, $r_p$, using $Kn_m \sim$1 to get a very rough lower bound and $r_* \sim r_a \sim r_0$, Equation (9) gives $Q_c \gg \sim 10\ x10^8 W$. This is

*well above* the largest heating rate (~$1.6x10^8$ W) in Erwin et al. (2013). Therefore, Pluto's atmosphere at the New Horizon encounter will be highly extended with an escape rate very close to the energy-limited rate, but the flow in the continuum region will be subsonic.

In order to further test Equation (10), we performed DSMC simulations with a distributed heating model in which, for simplicity, we used Beer's law along the radial direction (Murray-Clay et al. 2009): $q_a(r) = \varepsilon\sigma_a n(r) F_{UV/EUV} \exp(-\sigma_a N(r))$, where $\sigma_a$, $\varepsilon$, and $F_{UV/EUV}$ are the absorption cross section, the heating efficiency and solar energy flux at the upper boundary of our domain, $r = r_u$, and $N(r) = \int_r^{r_u} n(\bar{r})d\bar{r}$, with constant $\varepsilon$ out to $r_u$. The transition to a transonic solution with $r_*$ in the continuum region of the atmosphere was found to occur at $0.69 < Q_{net}/Q_c < 1.6$, in agreement with our criterion $Q_c$ calculated from Equation (10).

The threshold for transonic flow can be related to the absorbed energy converted to heat, $Q_a$, by accounting for the non-adiabatic cooling processes, $Q_{cool}$: $Q_{net} \sim Q_a - Q_{cool}$ (Erwin et al. 2013, Figure 4b). If $Q_{cool}$ is small, writing $Q_a \sim \varepsilon\pi r_a^2 F_{UV/EUV}$, then $\Phi_{EL}$ in Equation (5) can be roughly scaled by the luminosity *if* the solutions are *subsonic* or do not significantly exceed $Q_c$. Because the atmospheric expansion affects $r_a$, iterative solutions can improve estimates of $Q_a$. Ignoring this, we note that for an early earth-like upper atmosphere dominated by N or O, a solar EUV flux more than 100 times the present would be required for the escaping gas to go sonic in the continuum regime using data from Tian et al. (2008) in Equation (10). It is, therefore, unlikely that escape from such an atmosphere on a super-earth in the habitable zone would have a sonic point in the continuum regime.

Lammer et al. (2013) calculated the escape rate from a hydrogen atom thermosphere due to XUV radiation on super-earths observed orbiting close to their star. They used Parker's upper boundary conditions to solve the continuum equations and then decided that blow-off occurred if $\lambda_x < 3/2$ (Öpik 1963). Rather than solving these equations and then deciding whether the sonic or Jeans conditions should have been used, Equation (10) can be used to estimate whether a sonic point might occur in the continuum regime for a given $Q_{net}$. For example, using $Kn_m \sim 1$ and $r_* \sim r_0$ we find that $Q_c \sim 0.6 \ x10^{13}$ W and $2.5x10^{13}$ W for Kepler11b and 11c respectively, using data in Table 1 of Lammer et al. (2013). These values can be compared to their heating rates obtained using $\varepsilon = 0.15$: $Q_{net} \sim 1.2$ and $0.3x10^{13}$ W respectively. For this $\varepsilon$, Kepler11b has a sonic point ($Q_{net} > Q_c$) in the continuum region, whereas Kepler11c does not ($Q_{net} < Q_c$). Therefore, $\Phi_{EL}$ in Equation (5) is most applicable to Kepler11c which requires kinetic, not sonic, boundary conditions to obtain for an accurate description of its upper atmosphere. Of course, increasing $\varepsilon$ or reducing the gravitational energy due to the tides can change this.

**Summary**

We have used results from DSMC simulations to show that the oft-used energy limited rate in Equation (5) for an isentropic expansion of a heated upper atmosphere is most reasonable for a subsonic expansion with a large escape rate. The accuracy depends on how well one estimates $Q_{net}$. Conversely, agreement with the energy limited escape rate *does not* imply that sonic boundary conditions are applicable to continuum models of thermal escape. In fact the simple approximation in Equation (5) becomes less good with increasing $Q_{net}$ above $Q_c$ as seen in Figure 2. Although the *size* of the escape rate might not be strongly dependent on whether sonic, Jeans, or kinetic boundary conditions are used, the upper atmosphere can be significantly affected as seen in Figure 3. Therefore, past applications Parker's (1964a,b) model have led to incorrect descriptions of the upper atmosphere when rapid escape occurs (Tucker and Johnson 2009; Tucker et al. 2012; Erwin et al 2013). Since the upper atmosphere structure affects the interaction of the escaping gas with the ambient plasma and with neighboring bodies, we have given an expression in Equation (10) to estimate when sonic boundary conditions are likely to be applicable in calculating the escape from and the expansion of the upper atmosphere of a planetary body.

**Acknowledgement**

We acknowledge support from NASA's Planetary Atmospheres Program

**Appendix**

When $Q_{net}$ is absorbed in a layer at $r_a$ and $u(r)$ is small below $r_a$, Equations (1-2) can be integrated using $\kappa = \kappa_0 (T/T_0)^\omega$ and assuming zero gas velocity below the heated layer:

$$Q_{net} = E^+ + \frac{4\pi\kappa_0 T_0 r_0}{\omega+1} \frac{r_a/r_0}{r_a/r_0 - 1}\left[\left(\frac{\lambda_0 r_0}{\lambda_a r_a}\right)^{\omega+1} - 1\right] + 4\pi r_0^2 n_a \sqrt{\frac{\gamma}{m\lambda_a}} Ma_a \left[\frac{r_a}{r_0} U(r_0)\right]^{3/2} \frac{r_0/r_a + 1}{2},$$

where $Q_{net}$ is lost by energy carried off by escaping molecules, $E^+$, by downward thermal conduction (second term), and by adiabatic cooling (third term). Simulations having significant escape rates indicate that the last term dominates; assuming $r_a/r_0 - 1 \ll 1$ we obtain Equation (9).